\documentclass[reprint,
superscriptaddress,
 amsmath,amssymb,prb
]{revtex4-1}
\usepackage{amsmath}
\usepackage{amssymb}
\usepackage{graphicx}
\usepackage{dcolumn}
\usepackage{bm}
\usepackage{hyperref}
\usepackage{xcolor}
\usepackage{listings}
\usepackage{physics}

\usepackage{soul,xcolor}

\begin{document}
\title{Observation of exceptional point in a PT broken non-Hermitian system simulated using a quantum circuit}
\author{Geng-Li Zhang}
\email{gengli.zhang@link.cuhk.edu.hk}
\affiliation{Department of Physics, The Chinese University of Hong Kong, Shatin, N. T., Hong Kong, China}
\affiliation{Center for Quantum Coherence, The Chinese University of Hong Kong, Shatin, N.T., Hong Kong, China}
\author{Di Liu}
\affiliation{Central Research Institute, Huawei Technologies, Shenzhen 518129, China}
\author{Xi-Ming Wang}
\affiliation{Department of Physics, Southern University of Science and Technology}
\affiliation{Shenzhen Institute for Quantum Science and Engineering, Southern University of Science and Technology，Shenzhen 518055, China}
\author{Man-Hong Yung}
\email{yung@sustech.edu.cn}
\affiliation{Department of Physics, Southern University of Science and Technology}
\affiliation{Shenzhen Institute for Quantum Science and Engineering, Southern University of Science and Technology，Shenzhen 518055, China}
\affiliation{Guangdong Provincial Key Laboratory of Quantum Science and Engineering, Shenzhen Institute for Quantum Science and Engineering, Southern University of Science and Technology，Shenzhen 518055, Guangdong，China}
\affiliation{Central Research Institute, Huawei Technologies, Shenzhen 518129, China}

\begin{abstract}
Exceptional points (EPs), the degeneracy point of non-Hermitian systems,
have recently attracted great attention after its ability to greatly enhance the sensitivity of micro-cavities is demonstrated experimentally. 
Unlike the usual degeneracies in Hermitian systems, at EPs, both the eigenenergies and eigenvectors coalesce.
Although some of the exotic properties and potential applications of EPs are explored, the range of EPs studied is largely limited by the experimental capability.
Some of the systems, e.g. with higher-order EPs, are hard to achieve with conventional simulations.
Here we propose an extendable method to simulate non-Hermitian systems on the quantum circuits, where a wide range of EPs can be studied.
The system is inherently parity-time (PT) broken due to the non-symmetric controlling effects
and post-selection.
Inspired by the quantum Zeno effect, the circuit structure grantees the success rate of the post-selection.
A sample circuit is implemented in a quantum programming framework, and the phase transition at EP is demonstrated.
Considering the scalable and flexible nature of quantum circuits, our model is capable of simulating large scale systems with higher-order EPs.
We believe this work may lead to broader applications of quantum computers and provide a tool to the studies for non-Hermitian systems and the associated EPs. 
\end{abstract}

\date{\today}
\maketitle

\section{Introduction}

Quantum computation is long believed to be faster than the classical counterpart for many tasks.
The advantages of the quantum computation in various applications, such as factoring and searching \cite{Nielsen2010, Shor1997, Grover2001}, have been shown theoretically years ago.
However, the quantum supremacy, or advantage~\cite{Harrow2017} is only experimentally achieved by Google on their Sycamore processor recently~\cite{Arute2019}.
These newly available devices attracts considerable attention.
Among all researches on such noisy intermediate quantum chips, the simulation of the quantum systems may be one of the most practical and promising applications~\cite{Buluta2009, Brown2010, Georgescu2014}.
Most existing simulations~\cite{Gavroglu2011, AttilaSzabo1996} are designed for Hermitian systems.
This could be a natural choice considering the energy conservation of physical systems.
However, it is common that a system may be entangled and exchange energy with the environment.
After tracing out the environment, the evolution of the system follows an effective non-Hermitian Hamiltonian (i.e. $H\neq H^\dagger$)~\cite{El-Ganainy2018, Heiss1999, Heiss2000, Bender2002, Moiseyev2011}.
Therefore, the simulation of physical systems should not be limited to Hermitian systems.

Due to the unique properties of the exceptional point (EP), the degeneracy points of the non-Hermitian Hamiltonian, and the parity-time ($\mathcal{PT}$) phase transition \cite{Heiss2004, Bender2005, Berry2004}, the non-Hermitian physics has also attracted intensive interest recently.
In contrast to the conventional level degeneracy, at the EPs, not only the eigenenergies but also the corresponding eigenstates merge to be identical (coalesce) \cite{Heiss2004, Berry2004}.
This coalescence leads to many distinctive phenomena around EPs, such as the $\epsilon^{1/n}$ dependence of the level-splitting on the $\epsilon$ perturbations around the $n$th order EP \cite{Peng2014} and some nontrivial topological properties in the complex plane \cite{Xu2016, Leykam2017}.
Such properties raised vast new topics in the study of quantum sensing and system control~ \cite{Wiersig2014, Chen2017}.
For instance, though with some doubts \cite{Langbein2018, lau_fundamental_2018, zhang_quantum_2019, Chen2019, wang_petermann-factor_2020}, the last theoretical research and experimental evidence suggest that, EPs may be utilized for dramatically improve the sensitivity of level-splitting detection~\cite{Wiersig2014, Chen2017, Hodaei2017, Zhao2018}.

After first demonstrated in microwave cavities \cite{Dembowski2001}, the non-Hermitian effects were also soon observed in optical microcavities \cite{Lee2009, Rueter2010}, atomic systems \cite{Hang2013, Zhang2016}, electronics \cite{Bender2013, Assawaworrarit2017}, acoustics \cite{Zhu2014, Popa2014}, transmon circuits \cite{naghiloo_quantum_2019} and most recently nitrogen-vacancy centers in diamonds \cite{Wu2019}. 
However, the power of the fast developing quantum computing is largely ignored in the study of simulating non-Hermitian systems and investigating EPs. Here we propose a realization of non-Hermitian system to study EPs using the quantum circuits, which is applicable to NISQ devices.
Similar to the heralded entanglement protocols \cite{Lee2014}, the effective non-Hermitian model is heralded by measuring the ancillas to the $\lvert 0 \rangle$ states. 
For demonstration, a single qubit non-Hermitian system is simulated, where a phase transition at EP is observed.
This simulation is implemented with Huawei HiQ~\cite{HiQ},
a quantum programming framework based on the open-source python package ProjectQ \cite{Steiger2018, Haener2018}.
It is straightforward to generalized the method to multi-qubit systems and higher-order EPs, where an example is shown in Appendix.
We expect that the quantum chips in the near future could outperform the classical simulators for large non-Hermitian systems. Once the quantum chips are ready, the code can be migrated to the real device with minor modifications.
We believe this work paves the way for simulating non-Hermitian physics and investigating EPs with quantum computers.

\begin{figure*}
	\centering
	\includegraphics[width=0.9\linewidth]{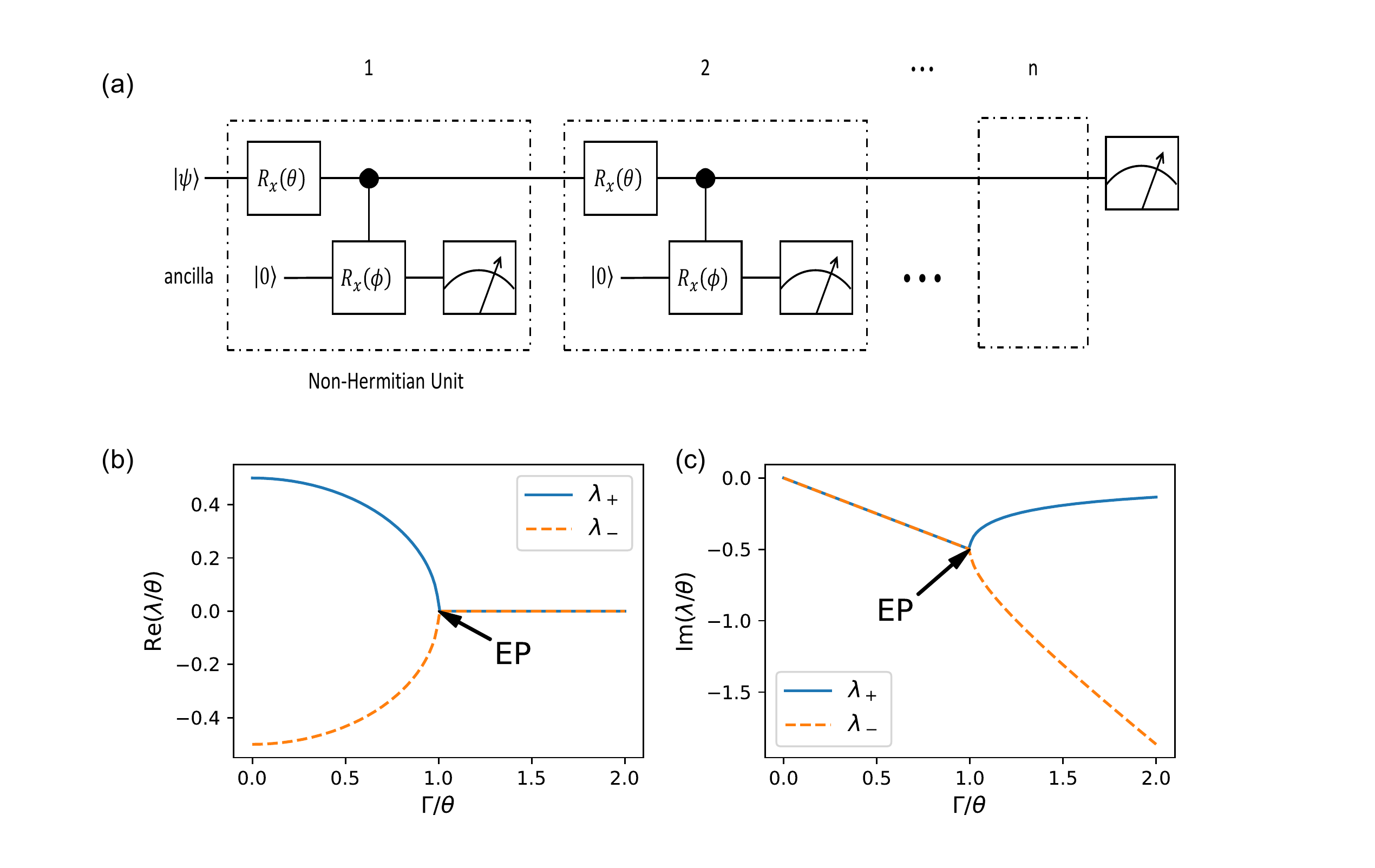}
	\caption{(a) The circuit for simulating a non-Hermitian system on quantum computers.
	$\lvert \psi\rangle$ is an arbitrary initial state of the system. For each cycle, the ancilla qubit is reinitialized to the $\lvert 0\rangle$ state.
	We only post-select the results with ancilla measured to be $\ket{0}$.
	(b)-(c) The real parts and the imaginary parts of eigenenergies of the effective non-Hermitian Hamiltonian.
	At the exceptional point (EP) both the eigenenergies and the eigenstates coalesce.
	The system is always in the PT-broken phase except for the point $\Gamma/\theta=0$. }
	\label{fig:circuitlambda}
\end{figure*}

\section{Model}

The motivation of our design of quantum circuit comes from the fact that, non-Hermiticity of physical systems are generated from the entanglement with the environments.
To imitate the real-world scenarios, the ``system'' qubits are entangled with the ancillas in the quantum circuit.
By measuring the ancilla qubits and post-selecting specific measurement results we can design the non-Hermiticity of the system qubits.

For simulations of two-dimensional non-Hermitian system, we take the circuit in Fig.~\ref{fig:circuitlambda}(a) as a concrete instance (see Appendix~\ref{sec: multi-qubits_circuit} for circuit that simulates higher dimensional system). The non-Hermiticity remains if the gates are replaced by other one- or two-qubits gates.
We take the first qubit as the ``system'' and the second qubit as an ancilla.
The non-Hermitian unit is repeated only if the measurement result of the ancilla is $\ket{0}$.
Since the measurement on ancilla is repeated in the same basis, similar to the quantum Zeno effect, the success rate can be boosted by dividing each unit to smaller units.
Starting from an initial state $\lvert \psi\rangle$, after $n$ cycles the final state of the system $\lvert \psi(n)\rangle$ is close to $\exp(-iH_{\mathrm{eff}}n)\lvert \psi\rangle$ ($\phi\ll 1$). 
The $H_{\mathrm{eff}}$ here is an effective non-Hermitian Hamiltonian (see Appendix.~\ref{Sec: Hamiltonian})
\begin{equation}\label{eq: non-Hermitian}
H_{\mathrm{eff}} = \frac{\theta}{2}\sigma_x + \frac{i\Gamma}{2}(\sigma_z - 1),
\end{equation}
where $\sigma_x$ and $\sigma_z$ are the Pauli operators, and $\Gamma = \phi^2/8$. This approximation is similar to Trotterization \cite{Lloyd1996, Yung2014}, whose error is $O(\Gamma^2)+O(\Gamma\theta)$.
The non-Hermiticity of the system comes from the post-selection on the ancilla qubit.
This process is similar to the non-Hermitian Hamiltonian in some quantum simulation experiments, such as
the one heralded by the absence of a spontaneous decay in cold-atom experiments \cite{Lee2014}.
It should be noted that the wavefunction evolved under the non-Hermitian Hamiltonian is unnormalized. It requires renormalization for further analysis.

The eigenenergies and the corresponding eigenstates of this Hamiltonian are
\begin{equation}
\lambda_\pm = -\frac{i\Gamma}{2} \pm \frac 12 \sqrt{\theta^2 - \Gamma^2}, \lvert v_\pm\rangle = \frac{1}{\mathcal{N}} \begin{bmatrix}
\frac 1\theta (i\Gamma \pm \sqrt{\theta^2 - \Gamma^2}) \\
1
\end{bmatrix},
\end{equation}
where $\mathcal{N}$ is a normalization constant. 

The real parts and the imaginary parts of the eigenergies are shown in Fig.~\ref{fig:circuitlambda}(b) and (c) respectively.
Since the imaginary part is nonzero except for the point $\Gamma/\theta=0$, this effective non-Hermitian system always lies in the PT-broken phase.
This is a result of the shared imaginary part $-\frac{i\Gamma}{2}$ in both eigenvalues.This term is essentially caused by the non-symmetric controlling effect (see Appendix.~\ref{Sec: Hamiltonian}).

\begin{figure*}[htb]
	\centering
	\includegraphics[width=0.85\linewidth]{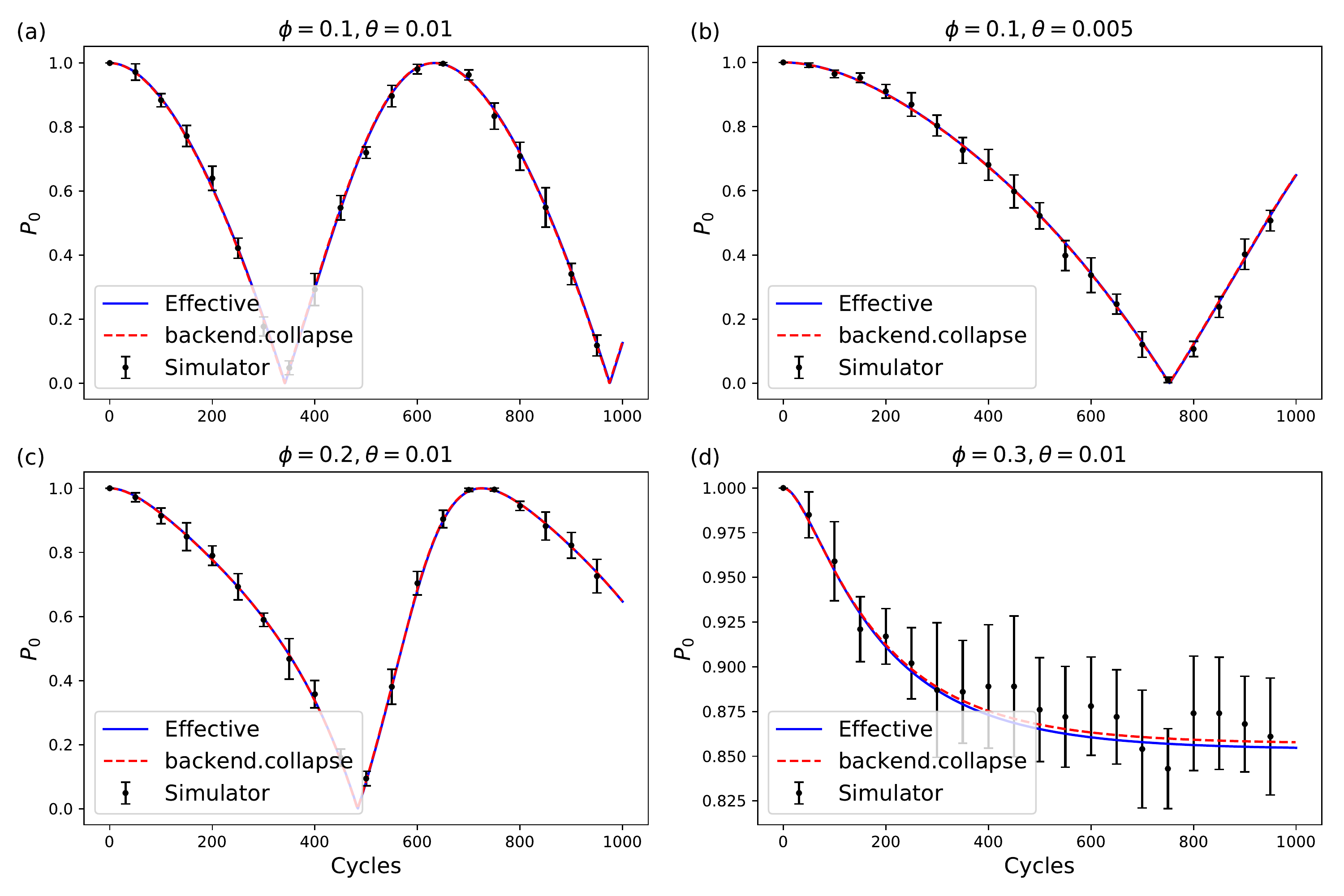}
	\caption[Compare effective and simulator]{Compare simulation results with theoretical results. (a-d) Results with different $\phi$ and $\theta$.
	The initial state of the qubit is set to $\lvert 0\rangle$.
	$P_0$ is the probability that the qubit remains in the $\lvert 0\rangle$ state.
	The blue lines are the theoretical results with the effective non-Hermitian Hamiltonian.
	The red dashed lines are the results using functions that are only available in the simulator backend as a quick verification of the circuits. The dots with error bars are the simulated results.
	The $n$-cycle non-Hermitian circuit is repeated $100$ times to get a single estimation of $P_0$.
	Then the whole process is repeated $10$ times to get the mean value of $P_0$ and the error bar. }
	\label{fig:compareeffsimcol}
\end{figure*}

\section{Implementation of the circuit}

The quantum circuit in Fig.~\ref{fig:circuitlambda}(a) can be implemented on any quantum devices that support the circuit-based quantum computing. To show that the simulated system is indeed non-Hermitian,
we implemented the circuit on the HiQ simulator. Once the quantum chips are ready and connected to the HiQ, we expect that the same algorithm can be run on the quantum chips with minor modifications (for instance, by changing the backend from simulator to quantum chips). 

The non-Hermitian unit in Fig.~\ref{fig:circuitlambda}(a) can be intuitively translated to the HiQ/ProjectQ language as:
\begin{lstlisting}[frame=single, language=Python, basicstyle=\ttfamily\footnotesize, commentstyle=\itshape\color{gray}]
# R_x rotation on the qubit
Rx(theta) | qubit  
# CR_x with qubit as control and ancilla as target
C(Rx(phi)) | (qubit, ancilla) 
# measurement on the ancilla
Measure | ancilla
\end{lstlisting}
where the standard Python \verb+|(or)+ operator is reloaded and used to apply the gates to qubits here.
\verb|qubit| and \verb|ancilla| are the qubits allocated in the \verb|MainEngine| representing the ``system'' and the ancilla respectively.
\verb|Rx(theta)| and \verb|C(Rx(phi))| are the rotational and the controlled rotational operators with respect to Pauli-X, and \verb|Measure| represents the quantum measurement in computational basis.

In order to implement the $n$-cycle non-Hermitian circuit in Fig.~\ref{fig:circuitlambda}(a), we first allocate a qubit and an ancilla. The qubit is initialized to an arbitrarily chosen state $\ket{\psi}$, and the ancilla is initialized to $\ket{0}$. After applying one non-Hermitian unit, it successes if the measurement result is $\ket{0}$.
If success, we allocate another ancilla which is also initialized to the $\lvert 0\rangle$ state, and repeat the first step. Otherwise, we start all over again.
The process is repeated until we achieve $n$ successes in a row, and the final state of the ``system'' should be proportional to $\ket{\psi(n)}$. In a trail, the whole process is repeated many times to estimate the probabilities $P_0 = \braket{0}{\psi(n)}$. Several trails are used to get the mean and the standard deviation.

Obtaining an accurate estimation by sampling is resource consuming.
In order to quickly verify if the circuit simulates an effective non-Hermitian system, we utilize the \verb|collapse_wavefunction| and \verb|cheat| functions that is only available to the simulator backend (see Appendix.~\ref{Sec: functions}). By using \verb|collapse_wavefunction(ancilla, [0])|, the post-selected wavefunction of the ``system'' with the ancilla at $\lvert 0\rangle$ is directly achieved.
Further, by using \verb|cheat()| the full information of the wavefunction can also be directly obtained. 

In Fig.~\ref{fig:compareeffsimcol}, the simulated result is compared to the analytical solution to the effective non-Hermitian Hamiltonian. The initial state is set to $\ket{\psi (0)}=\ket{0}$. It shows that, as long as $\phi\ll 1$ (important for both the Trotterization and the success rate as shown in Appendix.~\ref{Sec: Hamiltonian}), and  $\theta$ is small (so it is not far away from EP, see Sec.~\ref{Sec: EP_Phase}),
the circuit simulates the desired non-Hermitian system well. 

\begin{figure}[h!]
	\centering
	\includegraphics[width=0.95\linewidth]{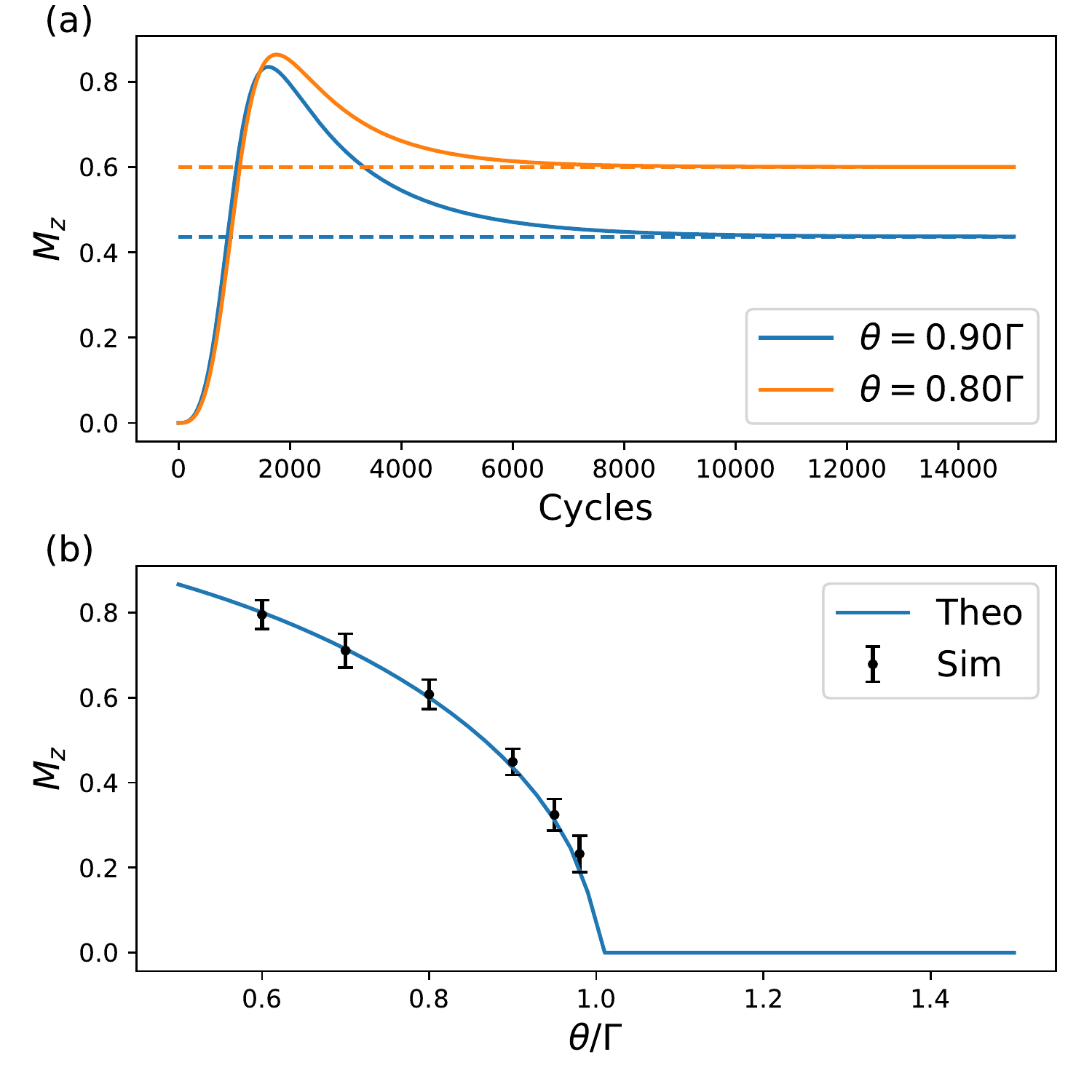}
	\caption[EPphase]{The stationary value and phase transiton at EP at $\phi=0.1$.
	(a) $M_z$ approaches a stationary value when $\Gamma > \theta$.
	The solid lines are results simulated with {\ttfamily collapse\_wavefunction} and {\ttfamily cheat} functions of the simulator backend. The dashed lines represents the theoretical stationary values.
	(b) shows the phase transitions of $M_z$ at EP.
	The $n$-cycle non-Hermitian circuit is repeated $500$ times for a single estimation of $M_z$, and the whole process is again repeated $20$ times to get the average and the 
	variance for each $\Gamma$.}
	\label{fig:mzstaturationepphase}
\end{figure}

\section{EP and phase transition}
\label{Sec: EP_Phase}

As marked in Fig.~\ref{fig:circuitlambda}(b), $\Gamma/\theta = 1$ is corresponding to the exceptional point (EP), where, unlike the Hermitian system, not only the eigenvalues but also the eigenvectors coalesce.
This degeneracy at EP leads to non-analytic behavior of the system \cite{Lee2014}, which can be easily observed by compute and plot $M_z \equiv \expval{\sigma_z}$ around EP.

When $\Gamma/\theta > 1$, the eigenvalues and eigenvectors can be recast as $\lambda_\pm = (-i\Gamma \pm i\theta \sinh\alpha)/2$ and $\lvert v_\pm \rangle = [i e^{\pm \alpha}, 1]^T$, with $\alpha = \cosh^{-1}(\Gamma/\theta)$.
In this regime, the eigenvalues have different imaginary parts, which means that the two eigenstates have different decay rate (though both negative) under time evolution.
The stationary state is $\lvert v_+ \rangle$ since its eigenvalue has a larger imaginary part (smaller absolute value), which implies $M_z = \langle v_+\rvert \sigma_z\lvert v_+\rangle = \sqrt{1-(\theta/\Gamma)^2}$.
As shown in Fig.~\ref{fig:mzstaturationepphase}(a), starting from the fully mixed state $\rho(0)=I/2$, $M_z$ gradually approaches the stationary value with the number of non-Hermitian cycles increases. 

When $\Gamma/\theta < 1$, we can rewrite the eigenvalues and the eigenvectors as
$\lambda_\pm = (-i\Gamma \pm \theta \cos\alpha)/2$ and $\lvert v_\pm \rangle = [\pm e^{\pm i\alpha}, 1]^T$ with $\alpha = \sin^{-1}(\Gamma/\theta)$. The two eigenvalues has the same imaginary parts, and therefore the two eigenvectors are equally stationary. As shown in Appendix.~\ref{Sec: Oscillating}, begin with the fully mixed state $\rho(0)=I/2$, $M_z$ will always oscillate in this regime.
For instance, when $\theta \gg \Gamma$ and at the long time limit, $M_z(t) \approx \sin(2\alpha)\sin(\theta t)$, from which it is not hard to see that, the lone time average vanishes, i.e., $M_z=0$. 

Therefore, the expectation value $M_z$ takes distinctive behaviors on each side of the EP point, which shows a phase transition even for a single qubit.
This is very different from the usual phase transition, which only happens when the number of particles goes to infinity. In Fig.~\ref{fig:mzstaturationepphase}(b) we show the good agreement between the simulations from HiQ and the theoretical results for $\Gamma>\theta$, which confirms this phase transition.
However, in the region of $\Gamma < \theta$, the huge number of cycles required for taking the time average, especially when $\Gamma$ is in the same order of $\theta$, is beyond our current scope. 

\section{Conclusion and discussion}

In summary, we've proposed a scheme of simulating non-Hermitian systems with quantum circuits, and numerically demonstrated the phase transition at EP of such system.
This is achieved by imitating the effect of environment with the post-selection of the measurement results on the ancilla qubits.
The codes for our numerical experiment is based on the simulator backend of HiQ, which can be cast to programs on physical quantum chips once they are available in the near future.
The non-Hermiticity of the quantum circuit have been shown and the phase transitions at EPs are also demonstrated.
Although the number of cycles ($> 5000$) required to show the phase transitions is hard to be achieved for quantum chips at this stage, the non-Hermiticity of the circuit  may be demonstrated experimentally on existing quantum chips ($\sim 100$ cycles).
Compared to previous implementations, which utilize the specific properties of the underlying systems, our method benefits from the universality and scalability of the quantum circuits. 
The idea of this work can be generalized to multi-qubit circuits and higher-order EPs, such as the one in Appendix~\ref{sec: multi-qubits_circuit}, where the advantage over other methods can be foreseen.
Our results could open a new path to the applications of quantum computers beyond the usual simulation paradigms that confined to Hermitian systems.

\begin{acknowledgments}
This work is supported by the Natural Science Foundation of Guangdong Province(Grant No.2017B030308003),the Key R\&D Program of Guangdong province (Grant No. 2018B030326001),the Science,Technology and Innovation Commission of Shenzhen Municipality(Grant No.JCYJ20170412152620376 and No.JCYJ20170817105046702 and No.KYTDPT20181011104202253),National Natural Science Foundation of China(Grant No.11875160 and No.U1801661),the Economy,Trade and Information Commission of Shenzhen Municipality(Grant No.201901161512), Guangdong Provincial Key Laboratory(Grant No.2019B121203002).
\end{acknowledgments}

\bibliography{references}

\clearpage
\appendix
\section{Non-Hermitian Hamiltonian after post-selection}
\label{Sec: Hamiltonian}

Consider the circuit in Fig.~\ref{fig:circuitlambda}(a).
For any intermediate state $\ket{\psi}$, 
Rewrite
the state after gate $R_x(\theta)$, 
$\exp(-i\theta \sigma_x/2)\lvert \psi\rangle\lvert 0\rangle$,
in computational basis:
$(\alpha\lvert 0\rangle + \beta \lvert 1\rangle)\lvert 0\rangle$.
Then the controlled rotation $CR_x(\phi)$ gives state
\begin{equation}
\xrightarrow{CR_x(\phi)} \qty(\alpha\lvert 0\rangle + \beta \cos\frac{\phi}{2}\lvert 1\rangle)\lvert 0 \rangle - i\beta \sin\frac{\phi}{2}\lvert 11\rangle.
\end{equation}

If the measurement result of the ancilla qubit is $\ket{0}$, assuming $\phi \ll 1$, the output should be
\begin{equation}
\begin{aligned}
\lvert \psi'\rangle &= \alpha\lvert 0\rangle + \beta (1-\phi^2/8+O(\phi^4))\lvert 1\rangle \\
&= e^{-i\frac{i\Gamma}{2}(\sigma_z - 1)+O(\Gamma^2)}(\alpha\lvert 0\rangle + \beta \lvert 1\rangle)\\
&= e^{-i\frac{i\Gamma}{2}(\sigma_z - 1)+O(\Gamma^2)}e^{-i\theta \sigma_x/2}\lvert \psi\rangle\\
&\approx e^{-iH_{\mathrm{eff}}}\lvert \psi\rangle, 
\end{aligned}
\end{equation}
where $\Gamma = \phi^2/8$.
the last step is similar to the Trotterization for quantum simulation with error $O(\Gamma^2)+O(\Gamma\theta)$, and the effective non-Hermitian Hamiltonian 
\begin{equation}
H_{\mathrm{eff}} = \frac{\theta}{2}\sigma_x + \frac{i\Gamma}{2}(\sigma_z - 1).
\end{equation} 

It should be noted that, the amplitude $\alpha$ 
is invariant under the non-Hermitian operation unit 
since the controlled rotation does not take effects when the qubit is at $\lvert 0\rangle$ state. This non-symmetry between the two eigenstates is the underlying reason that the system is always PT-broken. 

After the non-Hermitian unit, the probability 
of measuring $\ket{0}$ on the ancilla is $P_0 = 1-|\beta|^2\sin^2(\phi/2)$.
For instance, taken $\phi=0.1$ and on average $|\beta|^2=0.5$, we have $P_0\approx 0.999$.
After $5000$ cycles, the success rate is about $0.7\%$. 
However, since EP only depends on $\theta/\Gamma$, $\theta$ and $\Gamma$ can be scaled by a factor of $1/N$ at the same time without changing the EP.
Similar to the quantum Zeno effects, when $N$ approaches infinity, the total success rate approaches $1$, which can guarantee the observation of EPs as long as the fidelity of quantum chips is high enough. 

\section{Ciruit for multi-qubits}
\label{sec: multi-qubits_circuit}

\begin{figure}[h!]
    \centering
    \includegraphics[width=0.9\linewidth]{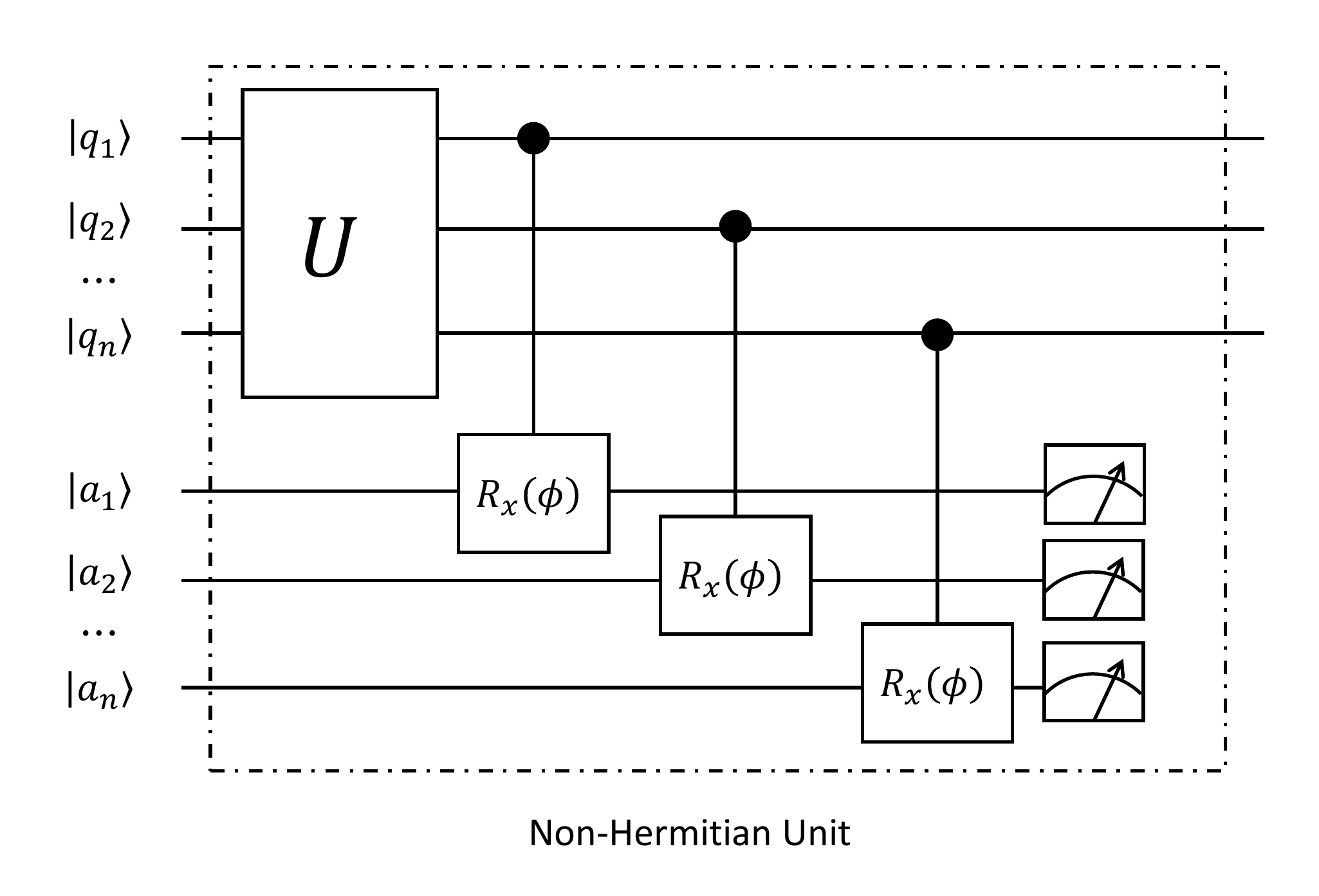}
    \caption{The circuit for simulating a $N$-qubits non-Hermitian system. The repeating strategy is the same as that has been described in the main text. $q_i$ denotes the $N$ system qubits, $a_i$ denotes the $N$ ancilla qubits and $U$ is the unitary evolution gates applied on the system qubits. The initial states of the system qubits can be arbitrary, but the initial states of the ancilla qubits are all $\lvert 0\rangle$ states.}
    \label{fig:N_circuit}
\end{figure}

With the similar idea of utilizing post-selection on the ansilla system, we can also simulate higher dimensional non-Hermitan system with the circuit structure shown in Fig.~\ref{fig:N_circuit}. 

Without the ancilla qubits, the system evolves under the unitary operator $U$ which corresponds to a Hermitian Hamiltonian $H$.
After taking into account of the ancilla qubits and post-selection on state $\ket{0\dots0}$ 
the effective Hamiltonian of the system reads
\begin{equation}
    H_{\mathrm{eff}} = H + \frac{i\Gamma}{2}\sum_i(\sigma_i^z - 1),
\end{equation}
where $\Gamma = \phi^2/8$ and $\sigma_i^z$ is the Pauli operator on the $i$th qubit.

With the effective multi-qubits non-Hermitian system, physics of higher order EPs can be investigated. For instance, assume the Hermitian Hamiltonian 
$H = a(\sigma^x_0 \sigma^z_1 + \sigma^y_0 \sigma^z_1) + b(\sigma^z_0 \sigma^x_1 + \sigma^z_0 \sigma^y_1)$and without loss of generality take $\Gamma=1$, we have four eigenenergies
\begin{equation}
    E_{u,v} = \frac 12(-2i + \sqrt{2}u \sqrt{4(a^2+b^2)-1 + v\sqrt{8a^2-1}\sqrt{8b^2-1}}),
\end{equation}
where $u=\pm 1$ and $v=\pm 1$. 

\begin{figure}[h!]
    \centering
    \includegraphics[width=0.9\linewidth]{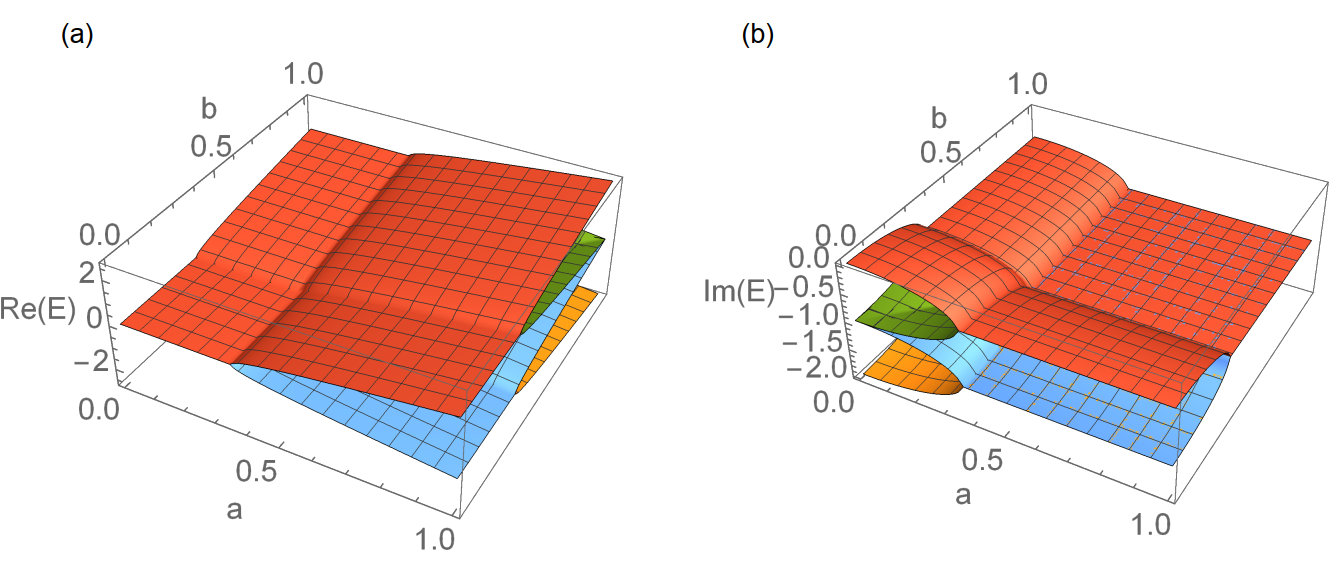}
    \caption{Eigenenergies of the two qubits non-Hermitian system. (a) The real parts of the eigenenergies. (b) The imaginary parts of the eigenenergies. Both the real and imaginary parts 
    degenerate at $a=b= 1/2\sqrt{2}$. }
    \label{fig:EP-4}
\end{figure}

When $a=b= 1/2\sqrt{2}$, the four eigenenergies coalesce, as shown in the Fig.~\ref{fig:EP-4}, and we have a $4$th order EP. 

\section{Useful functions of the simulator backend}
\label{Sec: functions}

Several functions in projectQ/HiQ are utilized in the simulator backend, which simplified the simulation of the circuit. But, to be noticed, they are only available to the simulator backend and cannot be used for simulations on real quantum chips.

The \verb|cheat| function can be used to directly access and manipulate the full wavefunction.
This function returns a list of two elements.
\verb|cheat()[0]| 
is
the mapping of the qubits with the bitlocations, which may depend
on how the compiler is optimized.
\verb|cheat()[1]| is
the amplitudes of the wavefunction.
they are stored as a \verb|numpy| array of length $2^n$ with $n$ being the number of qubits. 

The \verb|set_wavefunction| function is used to set the qubits to a specific state. 
This can be used a debugging tool, for instance, to verify the correctness the non-Hermitian units.

The \verb|collapase_wavefunction| can be used to directly obtain the desired post-selected wavefunction by specifying a specific measurement outcome (unless the probability is $0$), e.g., $\ket{0}$. 

The following code fragment illustrates the usage of these functions:  

\begin{lstlisting}[frame=single, language=Python, basicstyle=\ttfamily\footnotesize, commentstyle=\itshape\color{gray}]
eng.flush()
# set the initial wavefunction
eng.backend.set_wavefunction(psi_qbv[0], qubit)
H | qubit
ancilla = eng.allocate_qubit()
C(Rx(phi)) | (qubit, ancilla)

# the wavefunction when the ancilla is at 0
eng.flush()
eng.backend.collapse_wavefunction(ancilla, [0])
# deallocate the ancilla
del ancilla
eng.flush()
# print the wavefunction
print(eng.backend.cheat()[1])
\end{lstlisting}
where the \verb|flush()| function is required to push all of above gates to the simulator and execute. 

\section{Oscillating for $\Gamma < \theta$}
\label{Sec: Oscillating}

When $\Gamma < \theta$, the eigenvalues and eigenvectors can be recast as
\begin{equation}
\lambda_\pm = -\frac{i\Gamma}{2} \pm \frac{\theta}{2}\cos\alpha, \quad \lvert v_\pm\rangle =\frac{1}{\sqrt{2}}\begin{bmatrix}
\pm e^{\pm i\alpha} \\
1
\end{bmatrix},
\end{equation}
where $\alpha = \sin^{-1}(\Gamma/\theta)$.
The computational states can be written in eigenbasis:
\begin{equation}
\left\{
\begin{aligned}
\lvert 0\rangle &= \frac{1}{\sqrt{2}\cos\alpha}\left(\lvert v_+\rangle - \lvert v_-\rangle\right), \\
\lvert 1\rangle &= \frac{\sqrt{2}}{1+e^{i2\alpha}}\left(\lvert v_+\rangle + e^{i2\alpha}\lvert v_-\rangle\right).
\end{aligned}
\right.
\end{equation}

Starting from the initial $\lvert 0\rangle$ state, the state after the evolution is 
\begin{equation}
\begin{aligned}
\lvert \psi^{(0)}(t)\rangle &= \frac{1}{\sqrt{2}\cos\alpha}\left(e^{-i\lambda_+ t}\lvert v_+\rangle - e^{-i\lambda_- t}\lvert v_-\rangle\right) \\
&= \frac{e^{-\Gamma t/2}}{\sqrt{2}\cos\alpha}\left(e^{-i\frac{\theta t}{2}\cos\alpha}\lvert v_+\rangle - e^{i\frac{\theta t}{2}\cos\alpha}\lvert v_-\rangle\right),
\end{aligned}
\end{equation}
from which we can get the expectation value
\begin{equation}
M^{(0)}_z(t) = \frac{\cos^2(\frac{\theta t}{2}\cos\alpha - \alpha) - \sin^2(\frac{\theta t}{2}\cos\alpha)}{\cos^2(\frac{\theta t}{2}\cos\alpha - \alpha) + \sin^2(\frac{\theta t}{2}\cos\alpha)}.
\end{equation}

Similarly, starting from the initial $\lvert 1\rangle$ state, the expectation value is
\begin{equation}
M^{(1)}_z(t) = \frac{\sin^2(\frac{\theta t}{2}\cos\alpha) - \cos^2(\frac{\theta t}{2}\cos\alpha + \alpha)}{\sin^2(\frac{\theta t}{2}\cos\alpha) + \cos^2(\frac{\theta t}{2}\cos\alpha + \alpha)}.
\end{equation}

\begin{figure}[h!]
	\centering
	\includegraphics[width=0.9\linewidth]{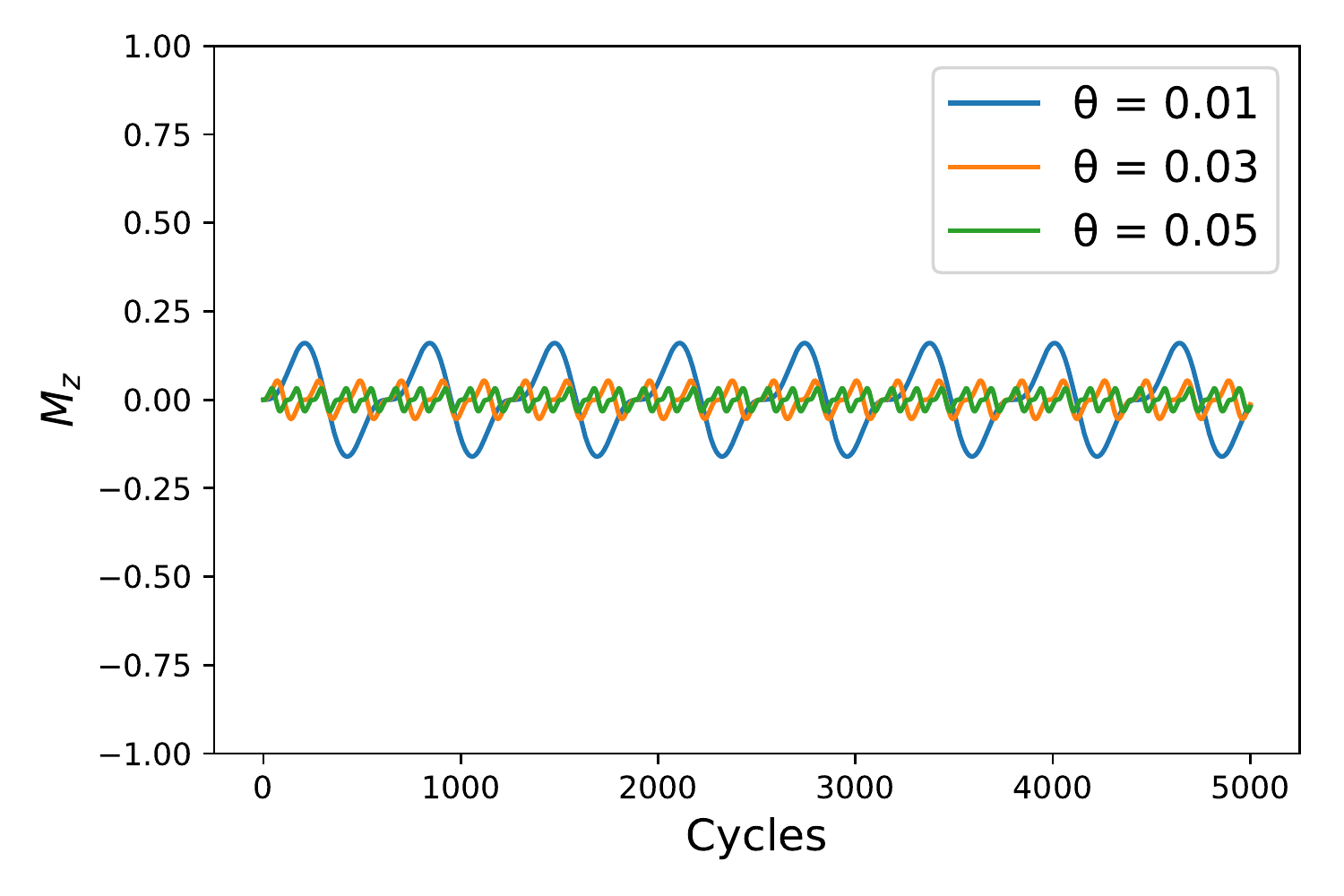}
	\caption[Oscillating Mz]{$M_z$ Oscillates for $\Gamma < \theta$, and $\phi$ is fixed at $0.1$.
	The results are simulated with {\ttfamily collapse\_wavefunction} and {\ttfamily cheat} functions of the simulator backend.The expectation values are always oscillating, and the long time average is vanishing.
	}
	\label{fig:mzoscillatingthetav}
\end{figure}

Therefore starting from the initial completely mixed state $\rho(0)=I/2$, the expectation value $M_z(t) = (M_z^{(0)}(t) + M_z^{(1)}(t))/2$ is always oscillating and hence has no stationary point. For instance, when $\theta \gg \Gamma$, $\alpha\approx \Gamma/\theta$, $\cos\alpha\approx 1-\Gamma^2/2\theta^2$, and at the long time limit ($t\gg 1/\theta$),
\begin{equation}
\begin{aligned}
M_z(t) &\approx \frac{1}{4}\left[\cos(\theta t\cos\alpha-2\alpha) - \cos(\theta t\cos\alpha+2\alpha)\right] \\
&=\frac{1}{2}\sin(\theta t\cos\alpha)\sin(2\alpha)\\
&\approx \frac{1}{2}\sin(2\alpha)\sin(\theta t),
\end{aligned}
\end{equation}
which means that the long time average of the expectation value is 0. 

\end{document}